\documentclass[a4paper]{llncs}
\usepackage{amsfonts,amsmath,amssymb,enumitem}
\usepackage{color}
\usepackage{alltt}
\usepackage{algorithm,algcompatible}
\usepackage{url}
\usepackage{graphicx}
\usepackage{epstopdf}
\usepackage{subfigure}
\urldef{\mailsa}\path|{flavio.ferrarotti, loredana.tec}@scch.at|
\urldef{\mailsb}\path|j.m.turull@massey.ac.nz|

\newtheorem{postulate}{Postulate}
\newcommand{\setofstates}[2]{\mathcal{#1}_#2}

\begin{document}

\title{Towards an ASM Thesis for Reflective Sequential Algorithms\thanks{The research reported in this paper results from the project \emph{Behavioural Theory and Logics for Distributed Adaptive Systems} supported by the {\bf Austrian Science Fund (FWF: [P26452-N15])}. The final publication is available at Springer via \url{http://dx.doi.org/10.1007/978-3-319-33600-8_16}}}

\author{}
\author{Flavio Ferrarotti\inst{1} \and Loredana Tec\inst{1}
 \and Jos\'{e} Mar\'{i}a Turull Torres\inst{2}
\institute{Software Competence Center Hagenberg, A-4232 Hagenberg, Austria\\
\mailsa\\
\and
Universidad Nacional de La Matanza, Argentina, and \\
Massey University, New Zealand \\
\mailsb\\
}
}
\maketitle

\begin{abstract}
Starting from Gurevich's thesis for sequential algorithms (the so-called ``sequential ASM thesis''), we propose a characterization of the behaviour of sequential algorithms enriched with reflection. That is, we present a set of postulates which we conjecture capture the fundamental properties of \emph{reflective sequential algorithms} (RSAs). Then we look at the plausibility of an ASM thesis for the class of RSAs, defining a model of abstract state machine (which we call \emph{reflective} ASM) that we conjecture captures the class of RSAs as defined by our postulates. 
\end{abstract}

\section{Reflective Sequential Algorithms}
\label{sec1}

In this paper we are concerned with linguistic reflection~\cite{[Stemple00]}, which can be defined as the ability of an algorithm to change itself.

In the field of computable functions this idea of reflection is as old as the field itself, think for instance of universal Turing machines. It has also been implemented in many programming languages. A prime example is LISP~\cite{[Smith84]}, where programs and data are represented uniformly as lists, and thus programs represented as data can be executed dynamically by means of an evaluation operator. Database theory is another field in which reflection has been deeply studied. It was shown that reflection can increase the expressive power of relational algebra \cite{[VandenBussche96]} and relational machines \cite{[Abiteboul98]}. Nowadays, most programming languages allow for some form of dynamic SQL, where the SQL queries are produced and evaluated dynamically during the program run-time, as opposed to static SQL where the queries are fixed at the time of compilation.

In the field of behavioural theory of algorithms however, linguistic reflection has not (up to our knowledge) been formally studied yet. This is surprising since dynamic self modifying code is a matter of increasing practical importance and key for the development of type-safe, dynamic agents, autonomous computing, and adaptive systems among others. The development of a good theoretical basis to describe, understand and prove properties of such systems, is then a pressing issue.   
 
Our aim in this work is to contribute to the development of a behavioural theory of reflective algorithms. In particular, we are concerned with \emph{reflective sequential algorithms} (RSAs), i.e., algorithms which are sequential in the precise sense of Gurevich's famous thesis~\cite{[Gurevich00]}, but which have the additional ability to change themselves. 

In the remaining part of this section we propose to capture the class of RSAs by means of three postulates which naturally extend the sequential time, abstract state and bounded exploration postulates in Gurevich's thesis~\cite{[Gurevich00]}. Then, in Section~\ref{ReflectiveASMs} we define a model of reflective ASM which we conjecture capture the class of RSAs as defined by our postulates. Section~\ref{examples} concludes this short paper with two examples of RSAs which satisfy our postulates. 

Similar to Gurevich's thesis for sequential algorithms~\cite{[Gurevich00]}, our first postulate states that every RSA works in sequential time. The key difference is that RSAs need to be able to change themselves. Thus, it seems natural to consider every state of a RSA as an \emph{extended state} which includes (a representation of) a sequential algorithm (in the precise sense of Gurevich's thesis~\cite{[Gurevich00]}) as part of it. In this way, transitions from one step to the next can also involve changes to the algorithm which now forms part of the state. Given a state $\bf S$ and a sequential algorithm $A$, we use $({\bf S},A)$ to denote an extended state which extends $\bf S$ with (a representation of) $A$. 

\begin{postulate}[Reflective Sequential Time Postulate]
\label{sequentialTime}
A RSA $\mathcal{A}$ consists of the following: 
\begin{itemize}
\item A non-empty set $\mathcal{S}_{\mathcal{A}}$ of \emph{extended states}, where each state is extended with (a representation of) a sequential algorithm which forms part of the state.
\item A non-empty subset $\mathcal{I}_{\mathcal{A}} \subseteq \mathcal{S}_{\mathcal{A}}$ of \emph{initial extended states} such that for all $({\bf S}_i, A_i), ({\bf S}_j, A_j) \in \mathcal{I}_{\mathcal{A}}$, $A_i = A_j$ (i.e., such that all initial extended states of $\mathcal{A}$ contain exactly the same sequential algorithm).  
\item A \emph{one-step transformation} function $\tau_{\mathcal{A}} : \setofstates{S}{\mathcal{A}} \rightarrow \setofstates{S}{\mathcal{A}}$ such that $\tau_{\mathcal{A}}(({\bf S}_i, A_i)) = ({\bf S}_j, A_j)$ iff $\tau_{A_i}(({\bf S}_i, A_i)) = ({\bf S}_j, A_j)$, where  $\tau_{A_i}$ denotes the one-step transformation function  of the sequential algorithm $A_i$. 
\end{itemize}
\end{postulate}
The concept of run remains the same as in the thesis for sequential algorithms, except that we consider extended states instead of arbitrary states. That is, a \emph{run} or \emph{computation} of $\mathcal{A}$ is a sequence of extended states $({\bf S}_0,A_0),({\bf S}_1,$
$A_1),$ $({\bf S}_2,A_2),\ldots,$ where $({\bf S}_0, A_0)$ is an initial extended state in $\mathcal{I}_A$ and $({\bf S}_{i+1},$
$A_{i+1}) = \tau_{\cal A}(({\bf S}_i,A_i))$ holds for every $i \geq 0$.  

While behavioural equivalent sequential algorithms have the same runs, this is not necessarily the case for RSA. In fact we can think of different runs $({\bf S}_0,A_0),({\bf S}_1,A_1),$ $({\bf S}_2,A_2),\ldots,$ and $({\bf S}_0',A_0'),({\bf S}_1',A_1'),$ $({\bf S}_2',A_2'),\ldots,$ where ${\bf S}_i = {\bf S}_i'$ and $A_i$ is behavioural equivalent (in the classical sense) to $A_i'$ for every $i \geq 0$. Since such runs clearly represent the same behaviour, we call them \emph{essentially equivalent runs} and define \emph{behavioural equivalent RSAs} as RSAs which have essentially equivalent classes of runs.   

As in the sequential ASM thesis, our second postulate defines (extended) states as first-order structures. However, extended states are not just arbitrary first-order structures, since each extended state must also include (an encoding of) a sequential algorithm given by a finite text. 
It is important to note that the vocabulary of a RSA is \emph{not} necessarily fixed. That is, we do not only allow RSAs to change themselves, but also to change their vocabularies. 
 
\begin{postulate}[Reflective Abstract State Postulate]
\label{abstractState}
\begin{itemize}
\item Extended states of RSAs are first-order structures.
\item Every extended state $({\bf S}, A)$ is formed by the union of an arbitrary first-order structure ${\bf S}$ and a \emph{finite} first-order structure ${\bf S}_A$ which encodes the sequential algorithm $A$.
% and maps every ground term used by $A$ to a different elements of the base set of $({\bf S},A)$ through a distinguished injective function.
\item The one-step transformation $\tau_{\mathcal{A}}$ of a RSA $\cal A$ does not change the base set of any extended state of $\cal A$. 
\item The sets $\setofstates{S}{\mathcal{A}}$ and $\setofstates{I}{\mathcal{A}}$ of, respectively, extended states and initial extended states of a RSA $\cal A$, are both closed under isomorphisms.
\item Any isomorphism between two extended states $({\bf S}_1,A_1)$ and $({\bf S}_2,A_2)$ of a RSA $\cal A$, is also an isomorphism from $\tau_{ \mathcal{A}}(({\bf S}_1,A_1))$ to $\tau_{\mathcal{A}}(({\bf S}_2,A_2))$. 
\end{itemize}
\end{postulate}

Our next (key) definition of strong coincidence of two extended states over a set of ground terms, is based on the fact that by the sequential accessibility principle of the ASM thesis for sequential algorithms~\cite{[Gurevich00]}, the only way in which $A$ can access an element $a$ of the base set of the state $({\bf S},A)$ is by producing a ground term that evaluates to $a$ in $({\bf S},A)$. 

\begin{definition}\label{def1}
Following the standard approach in reflective programming~\cite{[Stemple00]}, for every extended state $({\bf S}, A)$, we fix a total surjective function $\mathit{raise}_{({\bf S}, A)}: S_A \rightarrow \mathit{Ground}_A$ which maps (raises) elements from the domain $S_A$ of the finite structure ${\bf S}_A$ that encodes $A$, to (the level of) well formed ground terms in the finite set $\mathit{Ground}_A$ formed by all the ground terms used by the sequential algorithm $A$ to access elements of the extended state $({\bf S}, A)$. 

Let $({\bf S}, A)$ be an extended state, let $\Sigma_S$ and $\Sigma_A$ be the vocabularies of $\bf S$ and of the finite structure ${\bf S}_A$ which encodes the sequential algorithm $A$, respectively, let $\mathit{val}_{({\bf S},A)}(t)$ denote the interpretation in $({\bf S},A)$ of a ground term $t$ of vocabulary $\Sigma_S \cup \Sigma_A$, and let $\mathit{val}_{{\bf S}_A}(t)$ denote the interpretation in ${\bf S}_A$ of a ground term $t$ of vocabulary $\Sigma_A$. We say that two extended states $({\bf S}_1,A_1)$ and $({\bf S}_2,A_2)$ \emph{strongly coincide} on a set $W_S \cup W_A$ of ground terms of vocabulary $(\Sigma_{{\bf S}_1} \cup \Sigma_{A_1}) \cap (\Sigma_{{\bf S}_2} \cup \Sigma_{A_2})$ and $\Sigma_{A_1} \cap \Sigma_{A_2}$, respectively, iff the following holds:
\begin{itemize}
\item For every $t \in W_S$, $\mathit{val}_{({\bf S}_1,A_1)}(t) = \mathit{val}_{({\bf S}_2,A_2)}(t)$.   
\item For every $t \in W_A$ and corresponding $a_1 = \mathit{val}_{{\bf S}_{A_1}}(t)$ and $a_2 = \mathit{val}_{{\bf S}_{A_2}}(t)$,   
\begin{itemize}
\item $\mathit{raise}_{({\bf S}_1, A_1)}(a_1) = \mathit{raise}_{({\bf S}_2, A_2)}(a_2)$, and
\item $\mathit{val}_{({\bf S}_1,A_1)}(\mathit{raise}_{({\bf S}_1, A_1)}(a_1)) = \mathit{val}_{({\bf S}_2,A_2)}(\mathit{raise}_{({\bf S}_2, A_2)}(a_2))$.
\end{itemize}  
\end{itemize}
\end{definition}

We can now introduce our third and last postulate. It generalizes the bounded exploration postulate for sequential algorithms in~\cite{[Gurevich00]} to RSAs.
The key difference with the analogous postulate in the sequential ASM thesis, is that we use a stronger notion of coincidence. This is necessary because for each  RSA $\cal A$, we want to have a \emph{finite} bounded exploration witness set $W_{\cal A}$ which allows us to ``extract'' from every extended state $({\bf S}_i, A_i)$ of $\cal A$, a corresponding bounded exploration witness $W_{A_i}$ for $A_i$ (in the sense of the sequential ASM thesis).

 Let $({\bf S}, A)$ be the extended state of a RSA $\cal A$, we use $\Delta({\bf S},A)$ to denote the unique set of updates produced by the sequential algorithm $A$ in the extended state $({\bf S},A)$, which by virtue of Postulate~$1$ coincides with the set of updates produced by the RSA $\cal A$ in the extended state $({\bf S}, A)$. The formal definition of update and update set produced by a sequential algorithm is exactly the same as in~\cite{[Gurevich00],[BS03]}.

\begin{postulate}[Reflective Bounded Exploration Postulate]
For every RSA $\cal A$, there is a finite set $W_S \cup W_A$ of ground terms (called \emph{reflective bounded exploration witness}) such that $\Delta({\bf S}_1,A_1) = \Delta({\bf S}_2,A_2)$ whenever extended  states $({\bf S}_1, A_1)$ and $({\bf S}_2, A_2)$ of $\cal A$ strongly coincide on $W_S \cup W_A$.
\end{postulate}

A \emph{reflective sequential algorithm} (RSA) is an algorithm satisfying the Reflective Sequential Time, Reflective Abstract State and Reflective Bounded Exploration Postulates. In our next section we introduce a model of ASM machine which we conjecture characterizes this class of RSAs.  

\section{Reflective ASMs}\label{ReflectiveASMs}

The set of ASM rules of the reflective ASMs, as well as the interpretation of these rules in terms of update sets, coincide with those of the sequential ASMs as defined in~\cite{[Gurevich00]}. 

States of reflective ASMs are extended states. Each extended state $({\bf S}, R)$ of a reflective ASM is formed by the union of an arbitrary first-order structure $\bf S$ and a finite first-order structure ${\bf S}_R$ which encodes the sequential ASM rule $R$ as an abstract syntax tree $T_R$. ${\bf S}_R$ is formed by:
\begin{itemize} 
\item A finite set $V$ of nodes.
\item A finite set $L$ of labels which includes a different label for each ASM rule and each function symbol in the vocabulary of $({\bf S}, R)$.
\item A nullary function symbol $\mathit{self}$ interpreted as the root node of $T_R$.
\item Boolean binary function symbols $\mathit{child}$ and $\mathit{sibling}$ interpreted by the children and next sibling relationships of $T_R$, respectively.
\item A function symbol $\mathit{label}$ interpreted as a total labeling function of the nodes in $V$ with labels from $L$.
\item Nullary function symbols (constants) $l_{\mathrm{par}}$, $l_{\mathrm{if}}$, $l_{\mathrm{update}}$, $l_{\mathrm{import}}$ interpreted by the labels in $L$ corresponding to the ASM rules ${\bf par}$, $\bf if$, $\bf update$ and $\bf import$, respectively.
\item A different nullary function symbol (constant) $l_f$ for each function symbol $f$ in the vocabulary of $({\bf S}, R)$, interpreted by the label in $L$ corresponding to the function symbol $f$.
\item A different nullary function symbol (constant) $\mathit{node}_{w_v}$ for each node $v \in V \setminus \{\mathit{self}\}$ which is interpreted by $v$. Here $w_v$ is the word in the language defined by the grammar $P \rightarrow n \mid P.n$ where $n \in \mathbb{N}$, such that $w_v = n$ if $v$ is the $n$-th child of $\mathit{self}$ and $w_v = w_{v'}.n$ if $v$ is the $n$-th child of the node $v'$. For instance, if $v$ is the node in $T_R$ corresponding to the second child of the first child of $\mathit{self}$, then the constant $\mathit{node}_{1.2}$ is interpreted by $v$. 
\end{itemize}

Let $\Delta({\bf S},R)$ denote the set of updates yielded by a sequential ASM rule $R$ on an extended state $({\bf S}, R)$. Let $({\bf S},R) + \Delta({\bf S},R)$ be the extended state obtained by applying the updates in $\Delta({\bf S},R)$ to $({\bf S},R)$. A \emph{reflective ASM} $\mathcal{M}$ is formed by:
\begin{itemize}
\item A non-empty set $\mathcal{S}_{\mathcal{M}}$ of extended states which is closed under isomorphisms.
\item A non-empty subset $\mathcal{I}_{\mathcal{M}} \subseteq \mathcal{S}_{\mathcal{M}}$ of \emph{initial extended states} such that for all $({\bf S}_1, R_1), ({\bf S}_2, R_2) \in \mathcal{I}_{\mathcal{M}}$, $R_1 = R_2$.	
\item A transition function $\tau_{\mathcal{M}}$ over $\mathcal{S}_{\mathcal{M}}$ such that 
$\tau_{\mathcal{M}}(({\bf S},R)) = ({\bf S},R)+\Delta({\bf S},R)$ for every $({\bf S},R) \in \mathcal{S}_{\mathcal{M}}$.
\end{itemize}
A \emph{run} of a reflective sequential ASM is a finite or infinite sequence of extended states $({\bf S}_0,R_0),({\bf S}_1,R_1),({\bf S}_2,R_2), \ldots$, where $({\bf S}_0,R_0)$ is an initial extended state in $\mathcal{S}_{\mathcal{M}}$ and $({\bf S}_{i+1},R_{i+1})=\tau_{\mathcal{M}}(({\bf S}_i,R_i))$ holds for every  $i \ge 0$.

%We conjecture that reflective ASMs characterize RSAs as defined in Section~\ref{sec1}.

\section{Examples}\label{examples}
 
Let the sequential ASM rule in Figure~$1$ be the rule encoded in the initial states of a reflective ASM $\cal M$. It follows that every run of $\cal M$ produces an infinite sequence of extended states $({\bf S}_0, R_0)$, $({\bf S}_1,R_1), \ldots$ where each state $({\bf S}_{i+1}, R_{i+1})$ is obtained by updating the sub-tree rooted at $\mathit{node}_{1.2}$ of the syntax tree of $R_i$ so that it encodes the term $g+\underbrace{a + \ldots + a}_{(i+1)-\text{times}}$ instead of $g+\underbrace{a + \ldots + a}_{i-\text{times}}$, and by updating the location $(f,())$ with the value of the term $g+\underbrace{a + \ldots + a}_{i-\text{times}}$ in ${\bf S}_i$.        
\begin{figure}[h]
%\begin{minipage}{6cm}
\begin{algorithmic}[0]
\small
\STATE {\bf par}
      \STATE {\quad}$f := g$
      \STATE {\quad}$\mathit{child}(\mathit{node}_{1},\mathit{node}_{1.2}):= \mathtt{false}$
      \STATE {\quad}$\mathit{sibling}(\mathit{node}_{1.1},\mathit{node}_{1.2}):= \mathtt{false}$
      \STATE {\quad}{\bf import} $v_1$, $v_2$ {\bf do} {\bf par}
      \STATE {\quad}{\quad}{\quad} $l(v_1):=l_+$
      \STATE {\quad}{\quad}{\quad} $l(v_2):=l_a$
      \STATE {\quad}{\quad}{\quad} $\mathit{child}(\mathit{node}_{1},v_1):= \mathtt{true}$
      \STATE {\quad}{\quad}{\quad} $\mathit{child}(v_1,\mathit{node}_{1.2}):= \mathtt{true}$
	  \STATE {\quad}{\quad}{\quad} $\mathit{child}(v_1,v_2):= \mathtt{true}$
      \STATE {\quad}{\quad}{\quad} $\mathit{sibling}(\mathit{node}_{1.1},v_1):= \mathtt{true}$
      \STATE {\quad}{\quad}{\quad} $\mathit{sibling}(\mathit{node}_{1.2},v_2):= \mathtt{true}$
      \STATE {\quad}{\quad}{\bf endpar}
\STATE {\bf endpar}	
\end{algorithmic}
\label{fig:fig1}
\caption{Sequential ASM rule in the initial states of $\cal M$.}
%\end{minipage}
%\begin{minipage}{6cm}
%\includegraphics[scale=.5]{reflection.eps}
%\caption{Updates to syntax tree in Fig.3}
%\end{minipage}
%\end{figure}
%\begin{figure}
%\label{fig:fig2}
%\centering
%\includegraphics[scale=.5]{reflectivemachine.eps}
%\caption{Syntax tree of ASM rule in Figure~$1$}
\end{figure}

%Let $v_i$ be a node of the syntax tree $T_{R}$ encoded in an extended state $({\bf S}, R)$ of $\cal M$. We use  $T_{v_i}$ to denote the sub-tree of $T_{R}$ rooted at node $v_i$ and assume that $v_i$ is also the value in the extended state $({\bf S}, R)$ that corresponds to the string encoded by $T_{v_i}$. Thus, provided that $T_{v_i}$ encodes a well formed term of the vocabulary of $({\bf S}, R)$, we get that $\mathit{raise}(v_i)$ (see our definition of strong equivalence in Section~1) results in a well formed term. Under this assumptions,

If for every extended state $({\bf S}, R)$ of ${\cal M}$, we fix the function $\mathit{raise}_{({\bf S}, A)}: S_R \rightarrow \mathit{Ground}_R$ in Definition~\ref{def1} to be such that,  $\mathit{raise}_{({\bf S}, A)}(a_i) = t_i$ iff the sub-tree of $T_R$ rooted at $a_i$ ($T_R$ been the syntax tree encoded in ${\bf S}_R$) corresponds to the well formed ground term $t_i$ in the set $\mathit{Ground}_R$ of ground terms in $R$, and $\mathit{raise}_{({\bf S}, A)}(a_i) = \mathtt{undef}$ otherwise. Then it is clear that the set $W = W_S \cup W_A$ where $W_S = \{\mathtt{true}, \mathtt{false}, l_+, l_a\}$ and $W_A = \{\mathit{node}_{1.2}\}$, is a reflective bounded exploration witness for $\cal M$.   

As a second example, we consider the relational reflective machine (RRM) defined in~\cite{[Abiteboul98]} as a formal machine that computes only computable queries, as opposed to Turing machines. In RRMs the input (relational) database is stored in the so called relational store and it can be accessed only through first-order logic queries that are dynamic, i.e., are built during the execution of the machine. As part of its proof of completeness, such reflection power is used to find out the size of the domain of the database. To that end, the machine goes on building (and then executing) for each $n \geq 1$ the sentence $\exists x_1 \ldots x_n (\bigwedge_{1 \leq i \neq j \leq n} (x_i \neq x_j))$ until it becomes false, meaning that the previous value of $n$ is the wanted size. Note that the kind of reflection that the RRM uses is a bit different to the one we propose in this work. We could call it ``partial reflection'', since the sequence of actions performed in each transition, except for the queries to the relational store, never changes. We could then think in a different definition of the reflective ASM to represent partial reflection, as follows. Essentially, we only add to the sequential ASM a rule ${\bf eval} \, t$, which takes a term $t$ as its argument, and interpret its value as the root of the syntax tree of a sequential ASM rule (other than ${\bf eval}$) which is then executed.   

\bibliographystyle{splncs03}
\bibliography{thesis}
\end{document}